\documentclass[fleqn,usenatbib]{mnras}

\usepackage{newtxtext,newtxmath}

\usepackage[T1]{fontenc}

\DeclareRobustCommand{\VAN}[3]{#2}
\let\VANthebibliography\thebibliography
\def\thebibliography{\DeclareRobustCommand{\VAN}[3]{##3}\VANthebibliography}

\usepackage{array,multirow}
\usepackage{graphicx}
\usepackage{color}
\usepackage{booktabs}
\usepackage[nolist]{acronym}
\usepackage{amsmath}
\usepackage[]{natbib} 
\usepackage{hyperref}
\usepackage{mathtools,needspace,enumitem,etoolbox,microtype}
\usepackage{url}

\newcolumntype{M}[1]{>{\centering\arraybackslash}m{#1}}
\newcolumntype{N}{@{}m{0pt}@{}}

\def\apj{ApJ}
\def\apjs{ApJS}
\def\apjl{ApJ}

\def\mnras{MNRAS}

\def\nat{Nature}

\def\aap{A\&A}

\begin{document}

\title[Accretion efficiency and BHs]{Constraining accretion efficiency in massive binary stars with LIGO--Virgo black holes}

\author[Bouffanais et al.] {Yann Bouffanais$^{1,2}$\thanks{E-mail:bouffanais@pd.infn.it, yann.bouffanais@gmail.com},  Michela Mapelli$^{1,2,3}$\thanks{E-mail:michela.mapelli@unipd.it}, Filippo Santoliquido$^{1,2}$,  Nicola Giacobbo$^{1,2,4}$,  \newauthor{Giuliano Iorio$^{1,2,3}$, Guglielmo Costa$^{1,2,3}$} \\ 
$^{1}$Physics and Astronomy Department Galileo Galilei, University of Padova, Vicolo dell'Osservatorio 3, I--35122, Padova, Italy\\
$^{2}$INFN-Padova, Via Marzolo 8, I--35131 Padova, Italy\\
$^{3}$INAF-Osservatorio Astronomico di Padova, Vicolo dell'Osservatorio 5, I--35122, Padova, Italy\\
$^{4}$School of Physics and Astronomy, Institute for Gravitational Wave Astronomy, University of Birmingham, Birmingham, B15 2TT, UK
     }

\maketitle
\begin{abstract}
The growing sample of LIGO--Virgo black holes (BHs) opens new perspectives for the study of massive binary evolution. Here, we study the impact of mass accretion efficiency and common envelope on the properties of binary BH (BBH) mergers, by means of population synthesis simulations. We model mass accretion efficiency with the parameter $f_{\rm MT}\in[0.05,1]$, which represents the fraction of mass lost from the donor which is effectively accreted by the companion. Lower values of $f_{\rm MT}$ result in lower BBH merger rate densities and produce mass spectra skewed towards lower BH masses. Our hierarchical Bayesian analysis, applied to BBH mergers in the first and second gravitational wave transient catalogue,  yields zero support for values of $f_{\rm MT}\lesssim{}0.6$, with a lower boundary of the 99\% credible intervals equal to $f_{\rm MT}= 0.59$. This result holds for all the values of the common-envelope efficiency parameter we considered in this study $\alpha_{\rm CE} \in [1,10]$. This confirms that gravitational-wave data can be used to put constraints on several uncertain binary evolution processes.

\end{abstract}


\section{Introduction}

In 2015, the LIGO--Virgo collaboration (LVC) announced the first direct detection of gravitational waves (GWs) emitted by the merger of a binary black hole (BBH, \citealt{abbottGW150914,abbottastrophysics,abbottO1}). In the first and second observing runs (hereafter, O1 and O2), nine additional BBHs and one binary neutron star (BNS) mergers were detected \citep{abbottO2,abbottO2popandrate}.  In addition, \cite{zackay2019}, \cite{udall2019}, \cite{venumadhav2020} and \cite{nitz2020} claim several additional BBH candidates, based on an independent analysis of the O1 and O2 LVC data.

The release of GWTC-2 has drastically increased the number of detections with 39 new GW events detected during the first part of the third observing run (hereafter O3a, \citealt{gwtc2}). Among these new detections, a number of special events have been reported including the second BNS merger (GW190425, \citealt{abbottGW190425}), the first BBH with unequal mass components (GW190412, \citealt{abbottGW190412}), the most massive BBH merger ever observed (GW190521, \citealt{abbottGW190521,abbottGW190521astro}) and GW190814, whose secondary mass might be either the lightest black hole (BH) or the most massive neutron star (NS) ever observed \citep{abbottGW190814}. This upward trend in the number of detections should further continue on with the improvement of current GW detectors and the upcoming third generation of GW detectors \citep{kalogera2019}, Einstein Telescope \citep{punturo2010} and Cosmic Explorer \citep{reitze2019}. 

As a result of this increase in the number of detections, we are now capable of putting constraints not only on single sources parameters, but on the whole population of compact objects \citep{abbottO2popandrate,gwtc2_pop, roulet2019}. This gives us a unique opportunity to address some open questions on the astrophysics of these objects \citep{stevenson2017,mandel2018,gerosa2017,fishbach2017,fishbach2018,fishbach2020,bouffanais2019,callister2020}.

The formation of BBHs from massive stars is usually studied by means of population synthesis codes, which trace the evolution of a massive binary system from its formation to the possible merger of its compact remnants \citep[e.g.,][]{tutukov1973,bethe1998,portegieszwart1998,belczynski2002,belczynski2008,voss2003,podsiadlowski2004,belczynski2016,eldridge2016,stevenson2017,mapelli2017,mapelli2018,mapelli2019,giacobbo2018b,giacobbo2020,klencki2018,kruckow2018,vignagomez2018,eldridge2019,spera2019,tanikawa2020,belczynski2020}. Population-synthesis codes face the challenge of modelling several evolutionary stages that are still barely understood, such as common envelope \citep[e.g.,][]{webbink1984,dekool1990,ivanova2013,fragos2019} and mass transfer. In particular, mass transfer via Roche lobe overflow (RLO) is a complex process and has a tremendous impact on the properties of massive binaries \citep[e.g.,][]{eggleton2006}. Different assumptions on the stability criteria for RLO mass transfer  and on the onset of common envelope \citep[e.g.,][]{hjellming1987,soberman1997,murguia2017,macleod2020} translate into dramatic differences on the statistics of BBHs \citep[e.g.,][]{dominik2012,dominik2013,demink2016,mandel2016,marchant2016}. 

Another major uncertainty about RLO is represented by the fraction of mass lost from the donor star which is actually accreted by the companion (hereafter, $f_{\rm MT}$). Different codes  either assume nearly conservative mass transfer \citep[e.g.,][]{hurley2002}, or adopt a fiducial $f_{\rm MT}=0.5$ \citep[e.g.,][]{belczynski2002,belczynski2008} or prefer a highly non-conservative approach \citep[e.g.,][]{kruckow2018}. \cite{kruckow2018} test the behaviour of different values of $f_{\rm MT}$, concluding that low values of mass accretion efficiency ($f_{\rm MT}\sim{}0.25$) produce more realistic BNS masses. 

Here, we investigate the impact of mass accretion efficiency on the mass spectrum and on the merger rate of BBHs by means of our binary population synthesis code {\sc mobse} \citep{mapelli2017,giacobbo2018}. Exploiting Bayesian hierarchical analysis, we compare our simulated BBHs against LVC observations in O1, O2 and O3a \citep{abbottO2,abbottO2popandrate}. We find that models with  $f_{\rm MT}\leq{}0.6$ struggle to match LVC observations, if we assume that isolated binary evolution is the only formation channel for BBHs.


\section{Astrophysical model}

\subsection{Population synthesis}
\label{sec_MOBSE} 

{\sc mobse}\footnote{\url{https://demoblack.com/catalog-and-codes/}} is a customized and upgraded version of {\sc bse} \citep{hurley2000,hurley2002}, in which we included a new treatment for the evolution and the final fate of massive stars \citep{mapelli2017,giacobbo2018,giacobbo2018b}.

Mass loss by stellar winds in massive hot ($\ge{}12500$ K) stars is described as $\dot{M}\propto{}Z^\beta$, where
  \begin{equation}
    \beta=
    \begin{cases}
      0.85, & \text{if}\ \Gamma_{e} < 2/3 \\
      2.45-2.4\Gamma_{e}, & \text{if}\  2/3\leq{}\Gamma_{e} < 1\\
      0.05, & \text{if}\ \Gamma_{e} \geq{} 1
   \end{cases}
\end{equation}
In the above equation, $\Gamma_{e}$ is the electron-scattering Eddington ratio. 

The outcome of core-collapse supernovae (SNe) is modelled following \cite{fryer2012}. In particular, we adopt the delayed model, in which the explosion is launched $>500$ ms after bounce. This model does not produce any mass gap between 2 and 5 M$_\odot$. 
Stars with final carbon-oxygen mass $m_{\rm CO}\ge{}11$ M$_\odot$ collapse to BH directly. Following \cite{timmes1996} and \cite{zevin2020}, we compute neutrino mass loss for both NSs and BHs  as
\begin{equation}
    m_\nu{}=\min\left[\frac{\left(\sqrt{1+0.3\,{}m_{\rm bar}}-1\right)}{0.15},\,{}0.5\,{}{\rm M}_\odot\right],
\end{equation}  
where  $m_{\rm bar}$ is the baryonic mass of the compact object. 
The resulting gravitational mass of the compact object is $m_{\rm rem} = m_{\rm bar} - m_\nu{}$. This leads to a maximum BH mass of $\sim{}65-70$ M$_\odot$ at low metallicity. As already discussed in \cite{giacobbo2018b}, even if we form BHs with mass up to $\sim{}65$ M$_\odot$, only BHs with mass $\leq{}40$ M$_\odot$ merge within a Hubble time from isolated binary evolution, as an effect of mass transfer and common envelope in tight binary systems.
      
Prescriptions for pair instability SNe and pulsational pair instability are also implemented, as described in \cite{spera2017} and \cite{mapelli2020}. Our treatment for electron-capture SNe is described in \cite{giacobbo2019}. Natal kicks are implemented as in \cite{giacobbo2020}, assuming that $v_{\rm kick}\propto{}m_{\rm ej}\,{}m_{\rm rem}^{-1}$, where $m_{\rm ej}$ is the mass of the ejecta. This model allows us to reproduce both the proper motions of young pulsars in the Milky Way \citep{hobbs2005} and the merger rate inferred from LVC data \citep{abbottGW190425}.

In the original version of {\sc mobse}, mass transfer via RLO is described as in \cite{hurley2002}. This yields a nearly conservative mass transfer if the accretor is a non-degenerate star.  Here, we introduce  an alternative model in which 
the mass accretion rate ($\dot{m}_a$) is described as
  \begin{equation}\label{eq:MT}
    \dot{m}_a=\left\{
    \begin{array}{ll}
      f_{\rm MT}\,{}|\dot{m}_d| & \textrm{if non-degenerate accretor}\\ \\
      \min{(f_{\rm MT}\,{}|\dot{m}_d|,\dot{m}_{\rm Edd})} & \text{otherwise},
    \end{array}
    \right.
  \end{equation}
where $\dot{m}_d$ is the mass loss rate by the donor star, $\dot{m}_{\rm Edd}$ is the Eddington accretion rate and $f_{\rm MT}\in{}[0,\,{}1]$ is the accretion efficiency. We explore  20 values of $f_{\rm MT}$ ranging from 0.05 to 1 by steps of 0.05. Here, we do not allow for super-Eddington accretion onto a degenerate accretor. The impact of super-Eddington accretion is discussed in \cite{bavera2021}.

Other binary evolution processes such as wind mass transfer, tidal evolution, common envelope and GW energy loss are taken into account as described in \cite{hurley2002}. We explore 10 different values of the common envelope parameter $\alpha_{\rm CE}$ linearly spaced in the interval $[1,10]$. 
Small values of $\alpha_{\rm CE}$ mean that the binary must considerably shrink to eject the  envelope, while large values of $\alpha_{\rm CE}$ translate into easy ejection. According to the original definition of $\alpha_{\rm CE}$ as the fraction of orbital energy which is efficiently transferred to the envelope \citep{webbink1984}, values larger than one should be deemed unphysical. However, several studies have shown that additional sources of energy play a role during common envelope, which are not accounted for in the original $\alpha_{\rm CE}$ model \citep[e.g.,][]{ivanova2013,fragos2019}. Moreover, values of $\alpha_{\rm CE}\ge{}3$ seem to be in better agreement with the merger rate density of BNSs inferred from the LVC \citep{giacobbo2020}. Hence, in the following simulations, we adopt even values of $\alpha_{\rm CE}$ much larger than one.

We have considered 12 different stellar metallicities: $Z = 0.0002$, 0.0004, 0.0008, 0.0012, 0.0016, 0.002, 0.004, 0.006, 0.008, 0.012, 0.016, 0.02. For each run, we have simulated $10^7$ binaries per each metallicity comprised between $Z = 0.0002$ and 0.002, and $2\times{}10^7$ binaries per each metallicity $Z\ge{}0.004$, since higher metallicities are associated with lower BBH and BH--NS merger efficiency (e.g. \citealt{giacobbo2018b,klencki2018}). Thus, we have simulated $1.8\times{}10^8$ binaries per each value of $(f_{\rm MT},\alpha{}_{\rm CE})$.

%
%
%
             
The mass of the primary is randomly drawn from a Kroupa initial mass function \citep{kroupa2001} between 5 and 150 M$_\odot$. We derive the mass ratio $q=m_2/m_1$ as $\mathcal{F}(q) \propto q^{-0.1}$ with $q\in [0.1,\,{}1]$, the orbital period $P$ from $\mathcal{F}(\Pi) \propto \Pi^{-0.55}$ with $\Pi = \log_{10}(P/\text{day}) \in [0.15\,{}5.5]$ and the eccentricity $e$ from $\mathcal{F}(e) \propto e^{-0.42}~~\text{with}~~ 0\leq e \leq 0.9$ \citep{sana2012}.

\subsection{Merger rate density}
\label{sec_merger_rate}

To derive the merger rate density evolution of our models, we make use of the code {\sc cosmo$\mathcal{R}$ate} \citep{santoliquido2020a,santoliquido2020b}:
\begin{eqnarray}
\label{eq:mrd}
   \mathcal{R}(z) = \frac{{\rm d}~~~~~}{{\rm d}t_{\rm lb}(z)}\int_{z_{\rm max}}^{z}\psi(z')\,{}\frac{{\rm d}t_{\rm lb}(z')}{{\rm d}z'}\,{}{\rm d}z'\nonumber{}\\
   \times{}\,{}\int_{Z_{\rm min}}^{Z_{\rm max}}\eta(Z) \mathcal{F}(z',z, Z)\,{}{\rm d}Z,
\end{eqnarray}
where $t_{\rm lb}(z)$ is the look-back time at redshift $z$, $Z_{\rm min}$ and $Z_{\rm max}$ are the minimum and maximum metallicity, $\psi{}(z')$ is the cosmic SFR at redshift $z'$ from \cite{madau2017}, $\mathcal{F}(z',z,Z)$ is the fraction  of compact binaries that form at redshift $z'$ from stars with metallicity $Z$ and merge at redshift $z$, and $\eta(Z)$ is the merger efficiency, namely the ratio between the total number $\mathcal{N}_{\text{TOT}}(Z)$ of compact binaries (formed from a coeval population) that merge within an Hubble time ($t_{\rm H_{0}} \lesssim 14$ Gyr) and the total initial mass $M_\ast{}(Z)$ of the simulation with metallicity $Z$:
\begin{equation}
\label{eq:eta}
    \eta (Z) = f_{\rm bin}f_{\rm IMF}  \frac{\mathcal{N}_{\text{TOT}}(Z)}{M_\ast{}(Z)},
\end{equation}
where  $f_{\rm IMF}=0.285$ is a correction factor that takes into account that only stars with mass $m>5$ M$_\odot$ are simulated, while $f_{\rm bin}$ is the binary fraction, defined as the number of stars that are members of binary systems divided by the total number of stars in a population. Observations show that the binary fraction should depend on the stellar mass \citep{moe2017}. Here we assume $f_{\rm bin}=0.4$, which is the average value over all the considered stellar masses. We refer to \cite{santoliquido2020a} for more details on  {\sc cosmo$\mathcal{R}$ate}.

\subsection{Analytic description of the model}



For a given astrophysical model parametrized by $\lambda$, the population of merging BBHs is described as 
\begin{equation}
\dfrac{\text{d}N}{\text{d} \theta} (\lambda) = N(\lambda) \,{} p_{\lambda}(\theta),
\label{eq_pop_BBH}
\end{equation}
where $\theta$ are the parameters of the merging BBHs, $N(\lambda)$ is the total number of mergers predicted by the model and $p_{\lambda}$ is the probability distribution associated with the parameters of the merging BBHs. In our analysis, $\lambda=(f_{\rm MT},\,{}\alpha{}_{\rm CE})$.

In practice, as we know that the GW detectors will not detect sources that are further away than a given horizon redshift $z_{\text{h}}$, we restrict ourselves to sources with redshift comprised between 0 and $z_{\text{h}}$, such that the number of detectable mergers in our model is 
\begin{equation}
N(\lambda) = \int_{z=0}^{z_{\text{h}}} \mathcal{R}(z) \dfrac{\text{d} V_{c}}{\text{d}z} \dfrac{T_{\text{obs}}}{1+z} \text{d}z,
\end{equation}
where $\mathcal{R}(z)$ is the merger rate density,  $\text{d} V_{c} / \text{d}z$ is the comoving volume element and $T_{\rm obs}$ is the observation time considered in the analysis. As our analysis is focused on sources detected by the LVC during the first two observing runs, we take a conservative value of $z_{\text{h}} = 2$ for our horizon of BBHs.

To estimate our model distribution, $p_{\lambda}$($\theta$), we consider three parameters $\theta = \lbrace \mathcal{M},q,z \rbrace$, where $\mathcal{M}$ is the chirp mass and $q$ the mass ratio. We define $\mathcal{M}=(m_1\,{}m_2)^{3/5}\,{}(m_1+m_2)^{-1/5}$ and $q=m_2/m_1$, where $m_2\leq{}m_1$. We do not consider the spins, because we assume that, at first order, spins are not affected by the efficiency of mass transfer. Hence, all the models we consider have the same spin distribution.

In practice, the distributions $p_{\lambda}$($\theta$) were then estimated from a catalogue of $N_{\rm tot} = 50 000$ sources representative of our model. Each entry of the catalogue gives the value of $\theta = \lbrace \mathcal{M},q,z \rbrace$ for the source, and we used kernel density estimation to obtain an estimation of $p_{\lambda}$($\theta$).

To construct our catalogues, we make use of the merger rate density inferred from {\sc cosmo$\mathcal{R}$ate} to derive the expected distribution of sources between redshift 0 and $z_{\rm h}$. Then, the code {\sc cosmo$\mathcal{R}$ate} combines in good proportion the sources from the various catalogues of metallicities simulated with {\sc mobse}, such that we have the proper distribution of masses in the redshift intervals considered.

\begin{figure*}
	\includegraphics[width=17 cm]{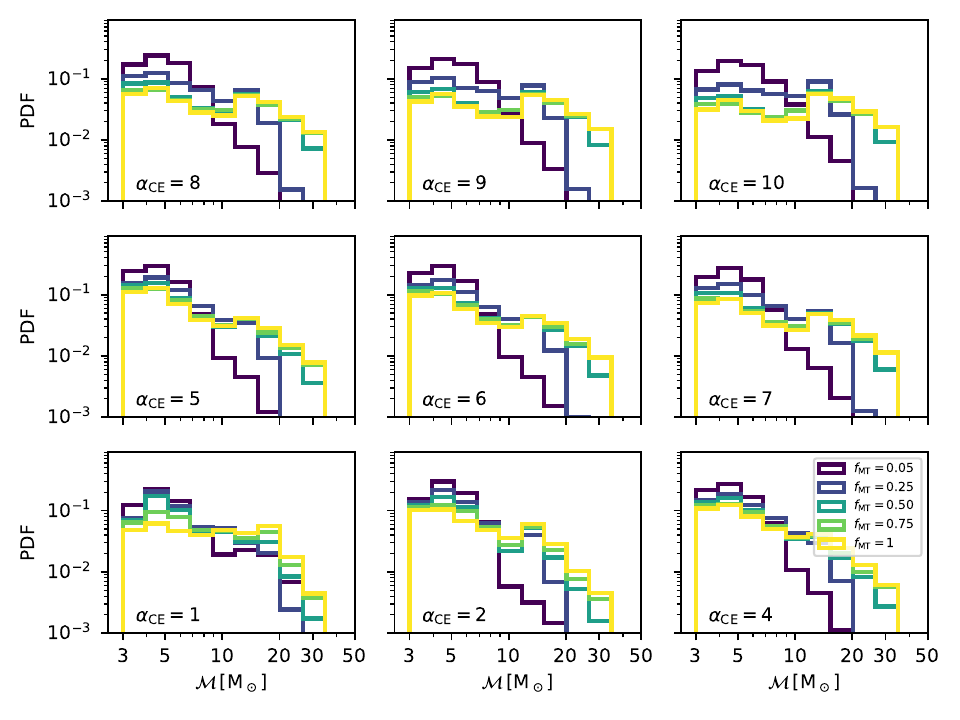}
    \caption{Chirp mass $\mathcal{M}$ distribution for a selection of our models. Upper row: $\alpha_{\rm CE}=8,$ 9, 10. Middle row: $\alpha_{\rm CE}=5,$ 6, 7. Lower row $\alpha_{\rm CE}=1,$ 2, 4. Different colors (from dark blue to yellow) refer to accretion efficiency $f_{\rm MT}=0.05$, 0.25, 0.5, 0.75 and 1.0. These distributions are integrated over redshift up to $z=2$.}
    \label{fig:mass}
\end{figure*}

\begin{figure*}
	\includegraphics[width=17 cm]{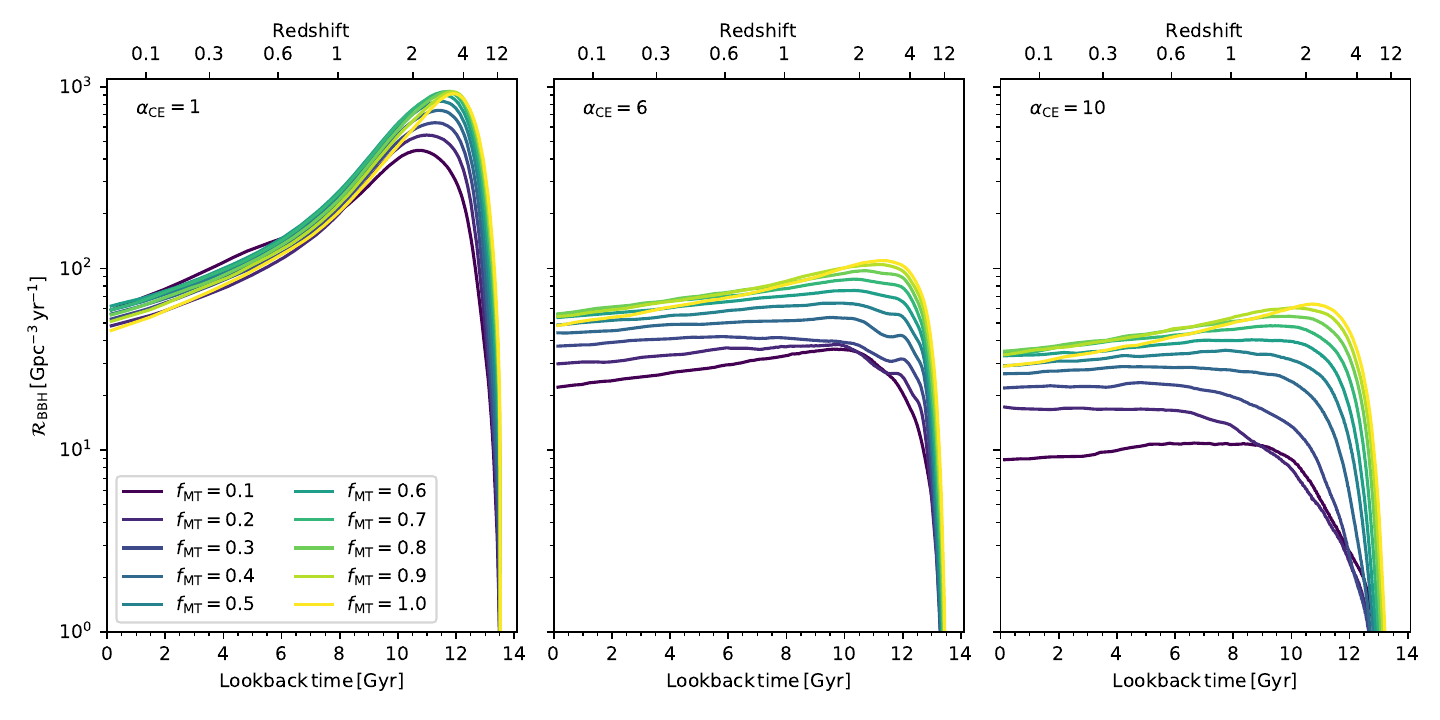}
    \caption{BBH merger rate density as a function of lookback time (lower $x-$axis) and redshift (upper $x-$axis). From left to right: $\alpha_{\rm CE}=1$, 6 and 10. Different colors (from dark blue to yellow) refer to accretion efficiency $f_{\rm MT}=0.1$, 0.2, 0.3, 0.4, 0.5, 0.6, 0.7, 0.8, 0.9 and 1.0.}
    \label{fig:MRD}
\end{figure*}

\section{Bayesian inference}

Bayesian analysis is widely used to analyze data from GW events. In particular, hierarchical Bayesian modeling has been the method of choice when doing model selection on population of compact binaries \citep{stevenson2015,fishbach2017,fishbach2018,stevenson2017,gerosa2017,vitale2017,zevin2017,talbot2017,talbot2018,mandel2018,taylor2018,abbottO2popandrate,fishbach2017,fishbach2018,wysocki2018,roulet2019,Kimball_2020,wong2019,vitale2019}.  In this section, we introduce the key concepts necessary to apply Bayesian analysis to a population of compact binaries. 

\subsection{Detector selection effects}

As our detectors are not perfect, the observed population can significantly differ from the original population, resulting in selection effects on the observed distribution. One way to quantify the response of the interferometer to an incoming GW signal is to compute the quantity $p_{\text{det}}(\theta)$, which represents the probability that a source with parameter $\theta$ is detected assuming one detector configuration \citep{finn1992,dominik2015}. By making use of this quantity, we can filter our original catalogue following the method used in \cite{bouffanais2019} and briefly summarized here.

For a given merging compact object signal, we can compute the value of its optimal signal to noise ratio (SNR), $\rho_{\text{opt}}$, corresponding to the case where the source is optimally oriented and located in the sky. All the uncertainties on the unknown position and orientation of the source can then be encoded in a parameter $\omega$, such that $\rho = \omega \,{} \rho_{\text{opt}}$ where $\rho$ is the actual SNR of the source and $\omega \in [0,1]$. The detection probability can then be expressed as a function of $\omega$ as,  
\begin{equation}
p_{\text{det}}(\omega) = 1 - F_{\omega} (\rho_{\text{thr}} / \rho_{\text{opt}}),
\end{equation}
where $F_{\omega}$ is the cumulative function of $\omega$ and $\rho_{\text{thr}}$ is a SNR threshold for which the value $\rho_{\text{thr}} = 8$ has been shown to be a good approximation of more complex analysis of detector network \citep{abadie2010,abbott2016rate,wysocki2018}. 
We use the fit provided by \cite{berti_refs} to compute an approximation of $F_{\omega}$ as a function of $\omega$.

The optimal SNR can be expressed as,
\begin{equation}
\rho_{\text{opt}} =4 \int \dfrac{\mid \tilde{h}(f) \mid^{2}}{S_{n}(f)} \text{d}f,
\end{equation}
where $\tilde{h}(f)$ is the waveform in the Fourier domain and $S_{n}(f)$ is the one-sided noise power spectral density. In our case, we have used the IMRPhenomD waveform \citep{khan2015} computed via the package {\sc PyCBC} \citep{canton2014,usman2015}. For the detector sensitivity, we have considered the sensitivities of O1, O2 and O3a separately. In practice, for each observing run we have computed the value of $S_{n}(f)$ by averaging the values of the one-sided noise power spectral density of the LIGO Livingston detector on all GW detections observed during a given observing run. This implies that we are slightly overestimating our network of detectors as we are assuming a network of Livingston-like detectors, while in reality the sensitivity of Hanford and Virgo were lower for all observing runs.

\subsection{Bayesian hierarchical modelling}

As our Bayesian hierarchical analysis framework has already been presented in \cite{bouffanais2019}, here we only summarize the main equations. For a given set of $N_{\rm obs}$ GW detections $\lbrace h \rbrace^{k}$, the posterior distribution of our model parameters $p(\lambda,N(\lambda) | \lbrace h \rbrace^{k})$ is associated with a likelihood corresponding to an inhomogeneous Poisson process \citep{loredo2004,mandel2018} 
\begin{eqnarray}
\mathcal{L}( \lbrace h \rbrace^{k} | \lambda )  \sim \text{e}^{-\mu(\lambda)} \prod_{k=1}^{N_{\rm obs}}  N(\lambda) \int \mathcal{L}^{k}(h^{k} | \theta)\,{} p_{\lambda}(\theta)\,{}  \text{d} \theta,
\label{post_hier_model} 
\end{eqnarray}
where $\mu(\lambda)$ is the predicted number of detections for the model and $\mathcal{L}^{k}(h^{k} | \theta)$ is the likelihood of the $k$th detection. The predicted number of detections is expressed as
\begin{equation}
\mu(\lambda) = N(\lambda) \,{} \beta(\lambda),
\end{equation}
where $\beta(\lambda)$ is the detection efficiency of the model defined as
\begin{equation}
\beta(\lambda) = \int p_{\lambda}(\theta)\,{} p_{\text{det}}(\theta) \,{}\text{d} \theta,
\label{efficiency_det}
\end{equation}
where $p_{\text{det}}(\theta)$ is the probability introduced in the previous section. 

From a computational point of view, the detection efficiency in eq.~\eqref{efficiency_det} is approximated with a Monte Carlo approach as
\begin{equation}
\beta(\lambda) \approx \dfrac{1}{N} \sum_{i}^{N}\,{} p_{\text{det}}(\theta_{i}),
\end{equation}
where the sum is evaluated over $N$ sources with parameters $\theta_{i}$. These parameters are drawn from the distribution $p_{\lambda}$ using a rejection sampling approach. 

For the integral with the GW-event likelihood, we also use a Monte Carlo method to approximate the expression such that we have
\begin{equation}
\int \mathcal{L}^{k}(h^{k} | \theta)\,{} p_{\lambda}(\theta)  \,{}\text{d} \theta \approx \dfrac{1}{N^{k}_{s}} \sum_{i=1}^{N^{k}_{s}} \dfrac{p_{\lambda}(\theta^{k}_{i})}{\pi^{\phantom{ }k}(\theta^{k}_{i}) },
\label{match_event}
\end{equation}
where $\theta^{k}_{i}$ is the $i$th sample of the $N^{k}_{s}$ samples from the posterior distribution of the $k$th GW detection provided by the LVC and $\pi^{\phantom{ }k}$ is the prior distribution for the parameters of the $k$th detection that we approximate using kernel density estimation.

\section{Results}\label{sec:results}

\subsection{Mass Spectrum}

Figure~\ref{fig:mass} shows the distribution of chirp masses of BBH mergers for a selection of our models. The distributions are integrated over redshift (up to $z=2$), since we do not observe a significant trend with redshift. This figure shows that 
BBH masses are affected by the accretion efficiency: larger values of $f_{\rm MT}$ lead to heavier BBH mergers. 
 A small value of $f_{\rm MT}$ implies that only a small fraction of the mass lost from the donor star during mass transfer is accreted by the companion. Hence, the total mass of the final BBH system will be lower than in the case of $f_{\rm MT}$ close to one. The shape of the mass spectrum is also affected by $\alpha_{\rm CE}$. In the Appendix~\ref{sec:appendix}, we discuss these features in detail.
 

\subsection{Merger Rate Density}

Figure~\ref{fig:MRD} shows the merger rate density evolution of a selection of our models. The value of $\alpha_{\rm CE}$ has a strong impact on the merger rate, as already discussed in \cite{santoliquido2020b}: small values of $\alpha_{\rm CE}$ lead to a higher local merger rate density of BBHs and to a steeper slope with redshift. The main reason is that smaller values of $\alpha_{\rm CE}$ are associated with shorter delay times between the formation and the merger of the BBH (Appendix~\ref{sec:appendix}). Hence, BBHs merge more efficiently if $\alpha_{\rm CE}$ is low and the peak of the merger rate is close to the peak of the metal-dependent star formation rate density. Actually, the peak of the merger rate density lies at higher redshift ($z\sim{}3-4$) than the one of the cosmic star formation rate density ($z\sim{}2$, \citealt{madau2017}), because the majority of BBH mergers are associated with metal-poor stars ($Z\leq{}0.002$), whose formation peaks at higher redshift. In contrast, larger values of $\alpha{}_{\rm CE}$ are associated with a lower merger efficiency and longer delay times (Appendix~\ref{sec:appendix}): a larger fraction of BBHs born at high redshift merge in the local Universe.

The Figure shows that the accretion efficiency also influences the merger rate: small values of $f_{\rm MT}$ lead to smaller BBH merger rate densities, especially for large values of $\alpha_{\rm CE}$. This happens because lower values of $f_{\rm MT}$ yield smaller BBHs, which are associated with longer coalescence timescales [$t_{\rm GW}\propto{}m_1^{-1}\,{}m_2^{-1}\,{}(m_1+m_2)^{-1}$, \citealt{peters1964}].
 
\subsection{Posterior distribution}

\begin{figure}
	\includegraphics[width=\columnwidth]{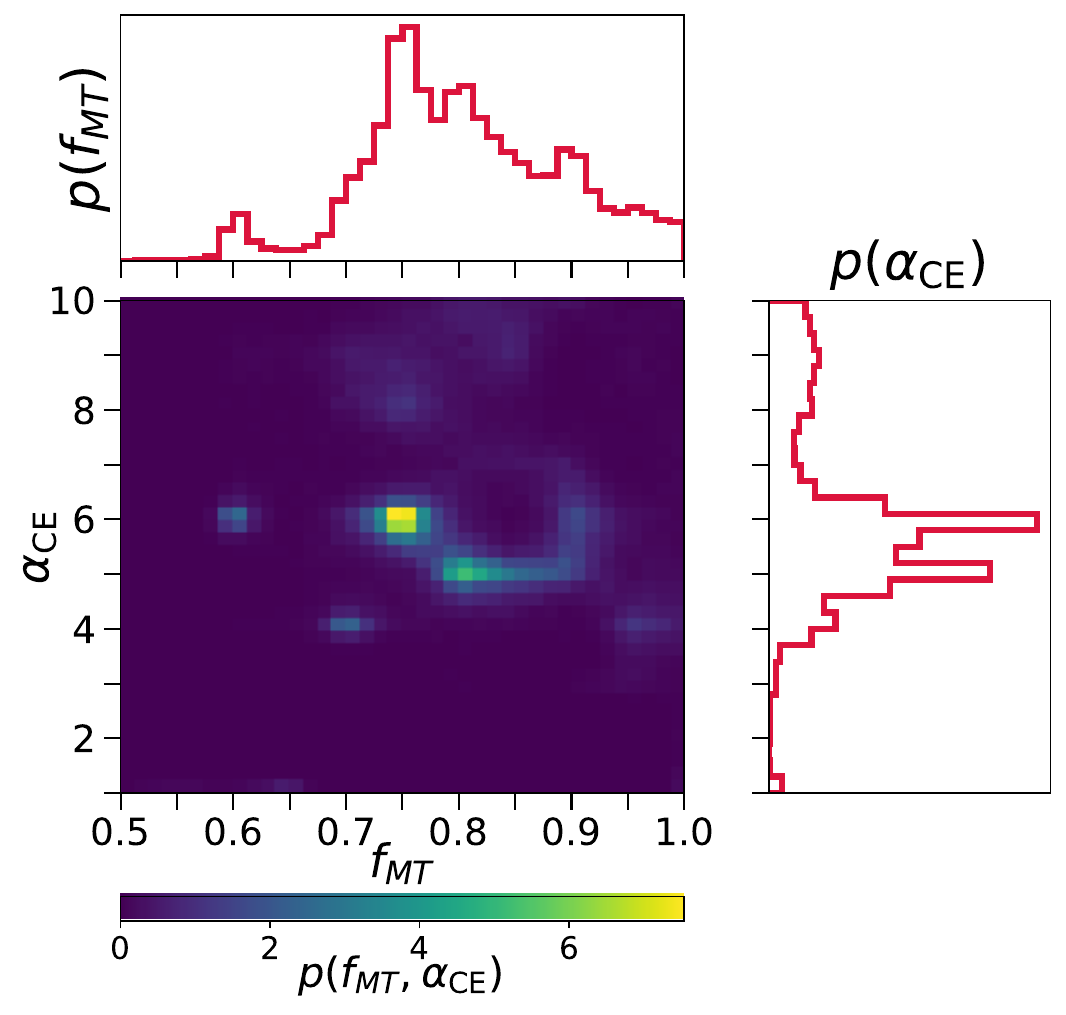}
    \caption{Posterior distribution of the parameters of our model $p(f_{\text{MT}},\,{}\alpha{}_{\rm CE})$ as inferred from O1, O2 and O3a excluding GW190521. The marginal posterior distributions $p(f_{\text{MT}})$ and $p(\alpha{}_{\rm CE})$ are given in the top and right panels respectively.}
    \label{fig:post_dist_fmt}
\end{figure}

Figure \ref{fig:post_dist_fmt} displays the posterior distribution of $p(f_{\text{MT}},\alpha{}_{\rm CE})$, as inferred from a MCMC chain produced via a Metropolis-Hastings algorithm run with the expression for the log-likelihood given in eq.~\eqref{post_hier_model} considering the GW events from 
GWTC-1 and GWTC-2, but excluding GW190521. Regarding the prior distribution, we assumed the following distribution, $\pi(f_{\text{MT}},\,{}\alpha{}_{\rm CE})=\pi(f_{\text{MT}})\,{}\pi(\alpha{}_{\rm CE})$, where $\pi(f_{\text{MT}})$ and $\pi(\alpha{}_{\rm CE})$ are two uniform distributions. The chain was run for $10^{7}$ iterations, after which we discarded the first $10^{4}$ iterations as burn-in and trimmed the chains using the information from the autocorrelation length. In addition, as we simulated only models on a discrete grid in the ($f_{\text{MT}}$,\,{}$\alpha{}_{\rm CE}$) space, we have used a bilinear interpolation for the logarithm of the likelihood for every targeted value of ($f_{\text{MT}}$,\,{}$\alpha{}_{\rm CE}$) explored by the MCMC algorithm. 

The most striking feature from the results in Figure \ref{fig:post_dist_fmt} is that the posterior distribution has zero support for values of $f_{\text{MT}} \leq 0.55$. A further inspection of the MCMC chain reveals that the lower boundary of the $99\%$ credible interval of the marginal distribution of $f_{\text{MT}}$ is equal to $0.59$.

While it is difficult to disentangle all the elements playing a role in this analysis, it is possible to interpret this result by looking at the mass spectra presented in Fig.~\ref{fig:mass}. In fact, smaller values of $f_{\text{MT}}$ result in BH mass spectra skewed to lower masses, that struggle to represent the most massive BBHs observed during the first three observing runs of LIGO--Virgo (like, GW150914 and GW170729).

Then, we observe that the maximum of the two-dimensional posterior distribution is around $(f_{\text{MT}},\,{}\alpha{}_{\rm CE})=(0.75,6)$. The shape of the posterior distribution is somewhat complex, and has support for several modes. Formulating a simple interpretation of the distribution shape is rather difficult, as it results from an interplay between the match of the model distribution with the observed events, the match of the model's rates with observed rates and corrections from selection effects.

Finally, both marginal distributions for $f_{\text{MT}}$ and $\alpha{}_{\rm CE}$ are relatively smooth over their support. The marginal distribution of $\alpha{}_{\rm CE}$ has support over the entire range of $\alpha{}_{\rm CE} \in [1,10]$, with a median of $5.81$, and the lower (upper) bound of the $99\%$ credible interval equal to $1.06$ ($9.93$). But we do observe some preferences for some model configurations, such as the double peaks at $\alpha{}_{\rm CE}=5$ and $6$.

As mentioned above, this analysis was done with the exclusion of the event GW190521. The reason 
is that all our models struggle to match such a high-mass event, with values of the integral in eq.~\eqref{match_event} of the order of $\sim10^{-146} $. Furthermore, some of our models, while still having extremely bad match with GW190521, yielded values for eq.~\eqref{match_event} that were orders of magnitude higher than the rest (i.e.  $\sim10^{-141}$). As a result, the latter models were drastically favoured by the analysis. A number of studies \citep[e.g.,][]{abbottGW190521astro,dicarlo2020a,dicarlo2020b,fragione2020quad,fragione2020b,rizzuto2020,fishbach2020, gayathri2020,deluca2020,liulai2021, Rice2020, romero-shaw2020,palmese2020,
 safarzadeh2020,samsing2020,kimball2020a,mapelli2021} have shown that GW190521 is much likely coming from another formation channel than isolated binary evolution. Thus, constraining the best isolated formation channel based on this event is intrinsically biased, which is why we decided to remove this event from the analysis. This also demonstrates the importance of having multiple formation channels in order to properly describe the entire set of events observed by LIGO--Virgo \citep{bouffanais2021,wong2021,zevin2021}.

\section{Conclusions}\label{sec:conclusions}

We studied the impact of mass accretion efficiency and common envelope on the properties of BBH mergers, by means of population synthesis (run with the code {\sc mobse}, \citealt{mapelli2017,giacobbo2018}) and Bayesian hierarchical analysis \citep{bouffanais2019}. In particular, we assume that, during a RLO mass transfer, a non-degenerate star can accrete only a fraction $f_{\rm MT}$ of the mass lost by the donor. We considered $f_{\rm MT}\in[0.05,1]$, and values of the common-envelope parameter $\alpha_{\rm CE} \in [1,10]$, and ran a large series of simulations (200) on a grid of ($f_{\rm MT}$,$\alpha{}_{\rm CE}$) with $\Delta f_{\rm MT}=0.05$ and $\Delta \alpha{}_{\rm CE} = 1$.

Models with low $f_{\rm MT}$ produce BBH mass functions skewed toward lower BH masses, because only a small fraction of the mass of the donor is accreted by the companion during the first Roche lobe mass transfer (Fig.~  \ref{fig:mass}). If $f_{\rm MT}\lesssim{}0.5$, the first born BH (i.e. the one that forms from the more massive component of the binary star) is also the more massive member of the BBH in $>50$\% of the simulated mergers. In contrast, for higher values of $f_{\rm MT}$, the second-born BH (which originates from the accretor of the first RLO episode) is  more massive than the first-born BH in most BBH mergers.

Models with low $f_{\rm MT}$ also yield smaller BBH merger rate densities (Fig.~\ref{fig:MRD}). The impact of $f_{\rm MT}$ on the merger rate is stronger for larger values of $\alpha_{\rm CE}$. If $\alpha_{\rm CE}=10$, the local BBH merger rate density spans from $\sim{}10$ Gpc$^{-3}$ yr$^{-1}$ if $f_{\rm MT}=0.05$ to $\gtrsim{}30$ Gpc$^{-3}$ yr$^{-1}$ if $f_{\rm MT}\gtrsim{}0.5$.

The common envelope parameter $\alpha_{\rm CE}$ has a more subtle impact on both mass and merger rate. Low values of $\alpha_{\rm CE}$ are associated with shorter delay times, because the common envelope phase is more efficient in shrinking the binary system. Hence, the merger rate has a steeper increase with redshift. In contrast, high values of $\alpha_{\rm CE}$ lead to longer delay times, resulting in a shallower increase of the merger rate with redshift. Also, larger vales of $\alpha_{\rm CE}$ tend to suppress the merger of light BHs, which have too long coalescence timescales.

We ran a hierarchical Bayesian analysis on our models against the LVC BBH mergers from O1, O2 and O3a \citep{abbottO2,gwtc2}. The posterior distributions (Fig.~\ref{fig:post_dist_fmt}) have almost zero support for values of $f_{\rm MT}\le{}0.6$. This result holds for all the values of $\alpha_{\rm CE}$ we considered in our study. Our models show a net preference for values of $\alpha_{\rm CE}\in[4,\,{}7]$ and $f_{\rm MT}\in[0.7,0.8]$.  
Models with low $f_{\rm MT}$ are strongly disfavoured because they yield lower merger rate densities and steeper BH mass functions, which struggle to represent the most massive BBHs in GWTC-2. 

In order to have these results, we had to ignore GW190521, the most massive BBH observed to date \citep{abbottGW190521,abbottGW190521astro}. The reason  is that all our models struggle to match such a high-mass event, with values of the integral in eq.~\eqref{match_event} of the order of $10^{-146}$. Here, we made the strong assumption that isolated binary evolution is the only channel to form BBH mergers. Dynamical formation channels are expected to produce heavier BBH mergers than isolated binary evolution (e.g., \citealt{mapelli2016,zevin2017,mckernan2018,dicarlo2019a,dicarlo2020a,bouffanais2019,rodriguez2019,antonini2019,mapelli2020b,rizzuto2020,fragione2020,arcasedda2020,banerjee2020}). Hence, assuming that  a fraction of the GWTC-2 BBHs have dynamical origin allows us to obtain a good match with GW190521 \citep{dicarlo2020b,mapelli2020b,mapelli2021} and might possibly reconcile a low accretion efficiency ($f_{\rm MT}<0.3$) with GW observations. In future studies, we will investigate the interplay between the ($f_{\rm MT}$, $\alpha_{\rm CE}$) parameters of binary evolution and the main dynamical formation channels. This poses a computational challenge but will give us a crucial key to interpret BBH formation.


\section*{Acknowledgement}
We thank the anonymous referee for their careful reading of the manuscript. We also thank Erika Korb, Michele Guadagnin, Alessandro Lambertini, Alice Pagano and Michele Puppin for useful discussions on mass transfer. MM, YB, FS, NG, GI and GC acknowledge financial support from the European Research Council for the ERC Consolidator grant DEMOBLACK, under contract no. 770017. 

\section*{Data availability}
The data underlying this article will be shared on reasonable request to the corresponding authors.

\appendix

\section{Mass and delay time}\label{sec:appendix}

\begin{figure*}
	\includegraphics[width=17 cm]{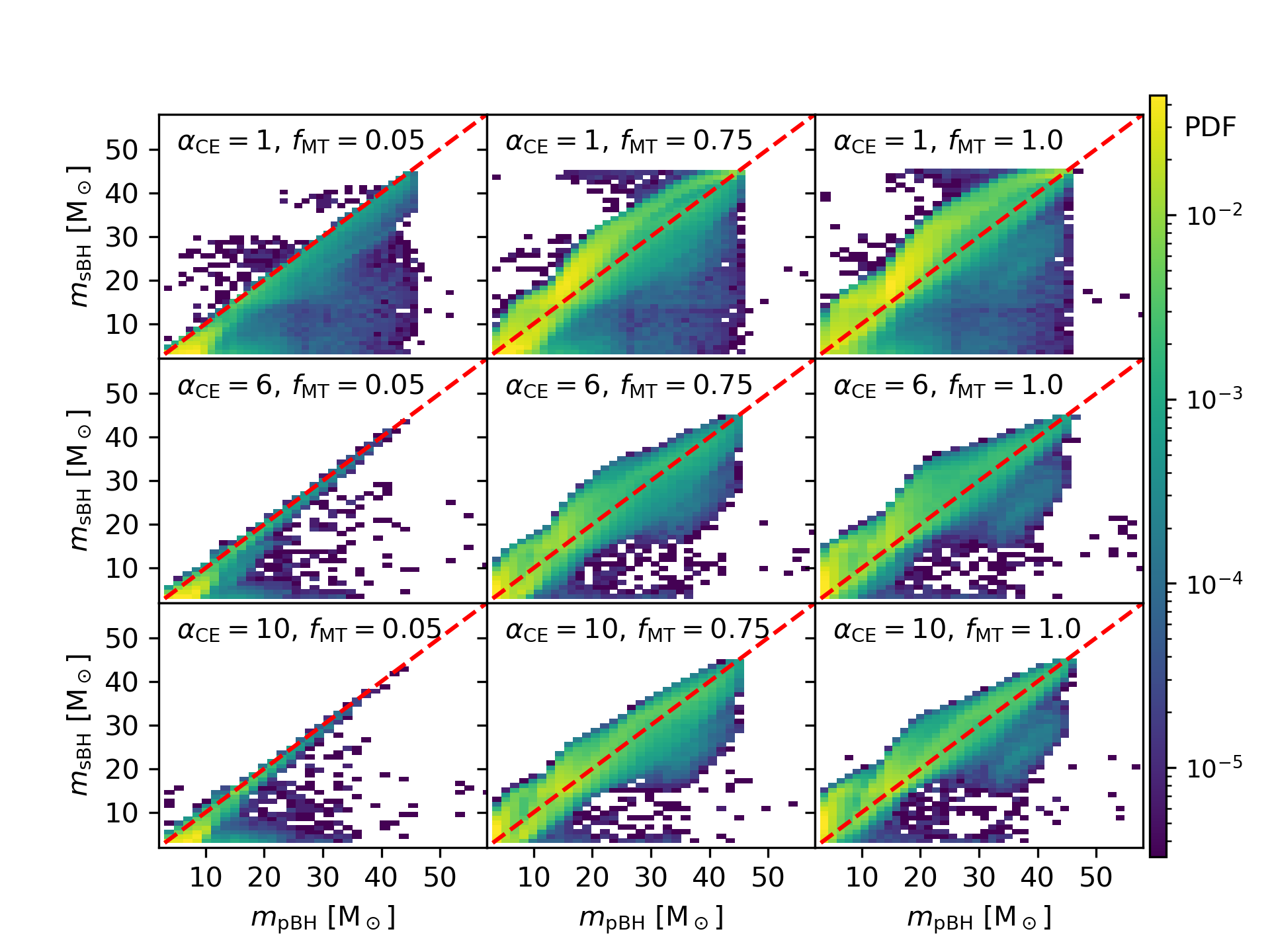}
    \caption{Mass of the secondary BH ($m_{\rm sBH}$) versus mass of the primary BH ($m_{\rm pBH}$) of all our BBH mergers. Here, the primary (secondary) BH is the one that forms from the most (least) massive component of the initial binary star. Hence, $m_{\rm pBH}$ can be less massive than $m_{\rm sBH}$. The BBHs shown in this Figure are not selected by merger redshift. Here, we show all the BBH mergers from the {\sc mobse} sample, stacking together our 12 progenitor metallicities.  Left-hand, middle and right-hand columns: $f_{\rm MT}=0.05,$ 0.75, 1.0. Upper, middle and lower rows: $\alpha_{\rm CE}=1,$ 6 and 10. The colour map shows the probability distribution function of each panel. The dashed red line is the identity line ($m_{\rm sBH}=m_{\rm pBH}$).}
    \label{fig:m1m2}
\end{figure*}

Figure~\ref{fig:m1m2} shows the mass of the secondary BH ($m_{\rm sBH}$) versus mass of the primary BH ($m_{\rm pBH}$) of all our BBH mergers. Here, the primary (secondary) BH is the one that forms from the more (less) massive component of the initial binary star. Hence, $m_{\rm pBH}$ can be less massive than $m_{\rm sBH}$. The BBHs shown in this Figure are not selected by merger redshift. Here, we show all the BBH mergers from the {\sc mobse} sample, stacking together our 12 progenitor metallicities. The progenitor metallicities are weighted by their merger efficiency.

The accretion efficiency has a major impact on the masses of the two BHs: a low value of $f_{\rm MT}$ is always associated with less massive BBHs than a high value of $f_{\rm MT}$, because a larger fraction of the initial stellar mass is lost from the system. For $f_{\rm MT}\sim{}1$, the more massive BH in the BBH often originates from the less massive progenitor, which accretes a significant fraction of mass from its companion. In contrast, if $f_{\rm MT}\sim{}0.05$, the more massive BH forms almost always from the more massive star in the original binary system: the less massive star does not accrete a significant fraction of its companion's mass and loses its envelope during the common envelope phase, yielding a rather small BH.

\begin{figure*}
	\includegraphics[width=17 cm]{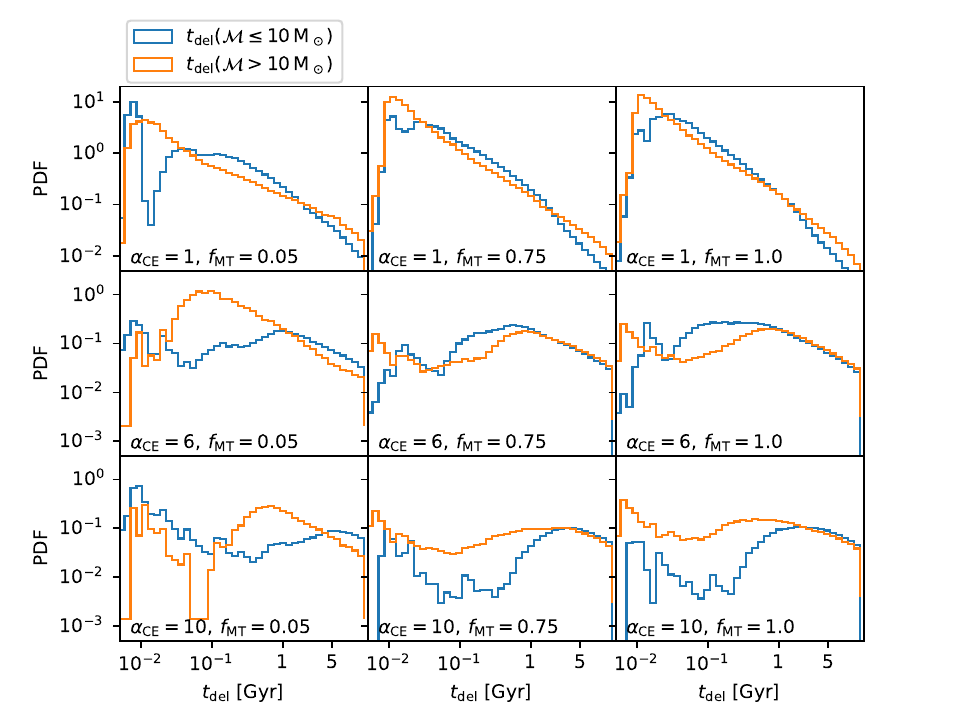}
    \caption{Distribution of the delay times $t_{\rm del}$ of our BBH mergers. Blue (orange) line: BBHs with chirp mass $\mathcal{M}\leq{}10$ M$_\odot$ ($\mathcal{M}>10$ M$_\odot$). The BBHs shown in this Figure are not selected by merger redshift. Here, we show all the BBH mergers from the {\sc mobse} sample, stacking together our 12 progenitor metallicities. The progenitor metallicities are weighted by their merger efficiency. Left-hand, middle and right-hand columns: $f_{\rm MT}=0.05,$ 0.75, 1.0. Upper, middle and lower rows: $\alpha_{\rm CE}=1,$ 6 and 10. }
    \label{fig:tdel}
\end{figure*}

Also, the common envelope efficiency $\alpha_{\rm CE}$ has a mild impact on the masses if we consider all the simulated BHs (Fig.~\ref{fig:m1m2}). The interpretation of this dependence on $\alpha_{\rm CE}$ is not as intuitive as that of $f_{\rm MT}$ and requires to introduce the concept of delay time ($t_{\rm del}$), i.e. the time elapsed between the formation of the progenitor binary star and the BBH merger. Figure~\ref{fig:tdel} shows the delay time of a selection of the simulated populations, distinguishing between BBHs with chirp mass $\mathcal{M}$ smaller and higher than 10 M$_\odot$.

A low value of $\alpha_{\rm CE}$ means that the binary system must shrink more in order to eject the envelope. Hence, after a common envelope phase, both low-mass and high-mass BBHs are able to merge with a relatively short delay time ($\lesssim{}0.2$ Gyr, Figure~\ref{fig:tdel}) if $\alpha_{\rm CE}$ is small.

For a large value of $\alpha_{\rm CE}$ ($\alpha_{\rm CE}\in{}[8,10]$), almost all binary progenitors survive the common envelope phase, but their semi-major axis does not shrink much during the process. Hence, if $\alpha_{\rm CE}\sim{}10$, more massive BBHs merge more likely than lighter BBHs, because they have shorter GW decay timescales [$t_{\rm GW}\propto{}m_{\rm sBH}^{-1}\,{}m_{\rm pBH}^{-1}\,{}(m_{\rm pBH}+m_{\rm sBH})^{-1}$, \citealt{peters1964}]. 

For intermediate values of $\alpha{}_{\rm CE}$ the situation is more difficult to predict, because it depends on the interplay between the delay time distribution (which in turn depends on the BBH mass) and the metallicity evolution of the Universe. This interplay yields the chirp mass distribution (up to redshift $z=2$) shown in Fig.~\ref{fig:mass}.

\bibliographystyle{mnras}
\bibliography{FMT_Project} 

\bsp	
\label{lastpage}
\end{document}